\documentclass[prd,aps,floats,tabularx,preprintnumbers,twocolumn]{revtex4}
\usepackage{graphicx, epsfig}
\usepackage{amsmath}
\usepackage{graphicx}
\usepackage{color}
\newcommand{\beq}{\begin{equation}}
\newcommand{\eeq}{\end{equation}}
\newcommand{\bea}{\begin{eqnarray}}
\newcommand{\eea}{\end{eqnarray}}

\def\gs{\mathrel{\raise1.16pt\hbox{$>$}\kern-7.0pt %
\lower3.06pt\hbox{{$\scriptstyle \sim$}}}}         %
\def\ls{\mathrel{\raise1.16pt\hbox{$<$}\kern-7.0pt %
\lower3.06pt\hbox{{$\scriptstyle \sim$}}}}         %

\begin{document}

\setlength{\unitlength}{1mm}
\title{Determining the Neutrino Mass Hierarchy with Cosmology}

\author{Francesco De Bernardis$^{1,2}$,
Thomas D.Kitching$^3$, Alan Heavens$^3$,Alessandro Melchiorri$^1$}
\affiliation{$^1$University of Roma ``La Sapienza``, P.le Aldo Moro
2, 00185, Rome Italy.\\$^2$Center for Cosmology, Dept. of Physics \&
Astronomy, University of California Irvine, Irvine, CA
92697.\\$^3$Scottish Universities Physics Alliance (SUPA), Institute
for Astronomy, University of Edinburgh, Blackford Hill, Edinburgh
EH9 3HJ,UK}

\bigskip

\begin{abstract}
The combination of current large scale structure and cosmic
microwave background (CMB) anisotropies data can place strong
constraints on the sum of the neutrino masses. Here we show that
future cosmic shear experiments, in combination with CMB
constraints, can provide the statistical accuracy required to answer
questions about differences in the mass of individual neutrino
species. Allowing for the possibility that masses are non-degenerate
we combine Fisher matrix forecasts for a weak lensing survey like
Euclid with those for the forthcoming Planck experiment. Under the
assumption that neutrino mass splitting is described by a normal
hierarchy we find that the combination Planck and Euclid will
possibly reach enough sensitivity to put a constraint on the mass of
a single species. Using a Bayesian evidence calculation we find that
such future experiments could provide strong evidence for either a
normal or an inverted neutrino hierachy. Finally we show that if a
particular neutrino hierachy is assumed then this could bias
cosmological parameter constraints, for example the dark energy
equation of state parameter, by $\gs 1 \sigma$, and the sum of
masses by $2.3\sigma$.
\end{abstract}


\maketitle

\section{Introduction}

Accurately determining the absolute value of the neutrino mass is one
of the main goals of particle physics. However since the effect of
individual neutrinos is small it is cosmological observations, that
observe the cumulative effect of neutrinos on large scales,
that present the most powerful way to bound the absolute neutrino mass
scale.
Albeit indirect and model dependent,
cosmological constraints are currently stronger than those
coming from beta-decay experiments (for recent reviews see
\cite{Fogli:2006yq} and \cite{Fogli:2008ig}). For example, the Mainz
\cite{Kraus:2004zw} and Troitsk \cite{Lob} Tritium decay experiments
give upper limits on the single electron neutrino mass of $m<2.05{\rm eV}$
and $m<2.3{\rm eV}$ respectively, at $95\%$c.l.. Cosmological data coming
from Cosmic Microwave Background (CMB) measurements from WMAP
\cite{Spergel:2006hy} experiment combined with baryonic oscillation
data \cite{Percival:2007yw} and various Supernovae observations
(\cite{Riess:2004nr}, \cite{Astier:2005qq}) lower this limit to
$m\lesssim0.22{\rm eV}$ \cite{Komatsu:2008hk}. On the other hand
observations of flavour oscillations in atmospheric and solar
neutrinos provide evidence not only for a non-zero neutrino mass, but
also for a difference between masses, measuring squared mass
differences of \cite{Maltoni:2004ei}:
\begin{equation}\label{sqmd}
\begin{split}
|\Delta m^{2}_{31}|=|m_3^2-m_1^2|=2.2^{+1.1}_{-0.8}\cdot10^{-3}{\rm eV}^2\\
\Delta m^{2}_{21}=m_2^2-m_1^2=7.9^{+1.0}_{-0.8}\cdot10^{-5}{\rm eV}^2
\end{split}
\end{equation}
where ranges indicated are $3\sigma$ confidence level and $m_1$, $m_2$
and $m_3$ are the three mass eigenstates. The ambiguity in the sign of
$\Delta m^2_{31}$ leads to an uncertainty about the neutrino mass
scheme, allowing for two possible hierarchies: the \textit{normal}
hierarchy, given by the scheme $m_3\gg m_2>m_1$, or the
\textit{inverted} hierarchy $m_2>m_1\gg m_3$. Note that given equation
($\ref{sqmd}$) an inverted hierarchy scenario would be automatically
excluded by measuring a total mass $\sum m_{\nu}\lesssim 0.1{\rm
  eV}$.

It is commonly perceived that cosmology is able to constrain the
total neutrino mass $\sum m_{\nu}$ while mass differences between
eigenstates can be neglected. This is an excellent approximation as
shown in \cite{Lesgourgues:2004ps} at least for the CMB anisotropies
power spectrum, since the effect of neutrino mass on the CMB is
related to the physical density of massive neutrinos $\omega_{\nu}$,
i.e. to their total mass $\sum m_{\nu}$.  Individual neutrino masses
do have an effect on the matter power spectrum, due to the different
transition redshifts from relativistic to non-relativistic
behaviour. This effect is still much smaller than that due to the
total mass and can be safely neglected in analysing current
cosmological data. Nevertheless in the near future various
experiments will reach a much higher accuracy in reconstructing the
matter power spectrum. It is therefore timely to consider the
possibility that these surveys will be sensitive to single neutrino
masses. In recent papers (see for example \cite{Hannestad:2002cn}
and \cite{Lesgourgues:2005yv}) a forecast has been made considering
future observations of CMB anisotropies, CMB lensing and galaxy
distribution finding that this kind of data doesn't seem able to
reach enough accuracy to discriminate between the two hierarchies.
In \cite{Pritchard:2008wy} has been shown that future accurate
measurements of the redshifted 21 cm signal from the epoch of
reionization can in principle measure individual masses but will be
very difficult to achieve the precision required to distinguish
normal and inverted hierarchies.

In \cite{Slosar:2006xb} an explicit and more general parameterization of
neutrino mass splitting was introduced,
representing various possible hierarchies;
with the only simplifying approximation that two neutrinos are of the same mass
$m_1=m_2$.  This approximation is well justified by equation
(\ref{sqmd}).
In \cite{Slosar:2006xb} it was shown that even a
cosmic variance limited CMB experiment would
not be able to detect a difference in individual neutrino masses. In
addition CMB
lensing, even if limited by cosmic variance only, is
strongly inhibited in measuring the mass hierarchy by degeneracies
with other cosmological parameters.

In this article we use a Fisher matrix formalism applied to the same
parameterization of \cite{Slosar:2006xb} to assess the ability of
future cosmic shear measurements, like those achievable with
Euclid/DUNE \cite{Refregier:2006vt} experiment, combined with Planck
\cite{:2006uk} CMB data to place constraints on single neutrino
masses. It is almost a decade that cosmic shear has been recognized
as one of the most powerful tools to constrain the total neutrino
mass (see for example \cite{Cooray:1999rv}, \cite{Hannestad:2006as}
and\cite{Kitching:2008dp}) and hence an investigation into how far
these experiments can proceed in the exploration of the neutrino
properties is well justified.

The article is
organized as follows: in section \ref{effects} we review in more
detail the effects of the total neutrino mass and of individual
neutrino masses on cosmological observables. We also describe the
parameterization of \cite{Slosar:2006xb} which will be used throughout
the rest of the article. In section \ref{forecast} we describe the
implementaton of our cosmic shear and CMB Fisher matrices. Section
\ref{results} shows results from our forecasts as function of various
assumptions about the neutrino mass hierarchy, and for different parameter
sets. We also discuss results in the light of our parameterization of
the neutrino mass hierarchy. Our forecasts show that accurate measurements
of the matter power spectrum from Euclid, combined with CMB data from
Planck, can reach the accuracy required to constrain single neutrino
masses. In section \ref{Bayes} we conduct a more accurate analysis of our
results through a Bayesian evidence investigation. We also show that
assuming a wrong hierarchy can lead
to a bias in the recovered value of other cosmological parameters, in
particular for the dark energy equation of state, generally comparable
with the $1\sigma$ sensitivity. The largest bias is found in the total
neutrino mass, due to the degeneracy involving $\sum m_{\nu}$ and
the hierarchy parameter, at $2.3\sigma$. Finally, in section \ref{conclusions} we
summarize our conlusions.

\section{Effects of neutrino mass splitting}\label{effects}

The main effect of a non-zero neutrino mass on cosmology is through
the collisionless fluid behaviour that causes neutrinos to
free-stream over a typical length which is known as
\textit{free-streaming length} $\lambda_{FS}$. The consequence of
this free propagation is a cancellation of neutrino density
fluctuations on scales smaller than $\lambda_{FS}$ and a slow-down in
the growth of perturbation on these scales. The matter power spectrum
results are then damped for wave vectors $k\gs k_{FS}\simeq 2\pi/\lambda_{FS}$. The free
streaming wave vector of a single species  $k_{FS}$ depends on the
mass of that species (\cite{Lesgourgues:2006nd})
\begin{equation}\label{kfs}
k_{FS}(z)=\frac{\sqrt{\frac{2}{3}[\Omega_m(1+z)^3+\Omega_{\Lambda}]}}{(1+z)^{2}}\left(\frac{m}{1{\rm eV}}\right)hMpc^{-1},
\end{equation}
where $h=H_0/(100 km\,s^{-1}\,Mpc^{-1})$. On these small scales the
matter power spectrum is suppressed with respect to the power
spectrum of a cosmology with massless neutrinos by an amount that
depends mainly on the fraction $f_{\nu}$ of matter density in the
form of massive neutrinos ($f_{\nu}\equiv \Omega_{\nu}/\Omega_{m}$).
In the case of degenerate masses $\Omega_{\nu}$ can be expressed as
a function of the total neutrino mass:$$\Omega_{\nu}\simeq\frac{\sum
m_{\nu}}{93.14h^2{\rm eV}}.$$ However, as pointed out in
\cite{Lesgourgues:2006nd}, even in the case of non-degenerate masses
this relation remains a good approximation.

Massive neutrinos become non-relativistic at a redshift given by
$z_{nr}\sim2\cdot10^{3}m_{\nu}/{\rm eV}$, so
that neutrinos with masses up to $\sim0.5{\rm eV}$ are still relativistic
at time of recombination. As a result the effect of neutrino free
streaming on the CMB power spectrum is negligible for small neutrino masses. In this case the
main effect of neutrino masses on the CMB is indirect, related
to the delay of matter-radiation equality.  This causes a small shift
in the peaks of the power spectrum and a slight increase of
their heights due to a longer duration of the Sachs-Wolfe effect.

From this discussion is clear that neutrino mass affects growth of
structure in two ways: the matter power spectrum is suppressed by an
amount that depends mainly on the total mass $\sum m_{\nu}$ but
also, even if for a minor amount, on single masses, because the time
of transition to non-relativistic regime depends on single masses.
The typical wave vector over which this suppression can be observed
also depends on the mass of single neutrino species as shown by
equation (\ref{kfs}). Therefore a reconstruction of the matter power
spectrum can in principle give information both on $\sum m_{\nu}$
and on single masses (through the suppression of the power spectrum
and $k_{FS}$), even if this second effect is generally much smaller
than that due to the total mass. In the following sections we
investigate the ability of future cosmological experiments to reach
the sensitivity required to detect differences in neutrino masses.
Following \cite{Slosar:2006xb} we parameterize mass splitting by
introducing the parameter $\alpha$ defined as the fraction of the
total mass in the third neutrino mass eigenstate:
\begin{equation}\label{alpha}
m_{3}=\alpha\sum m_{\nu}.
\end{equation}
The other two eigenstates are assumed to share the same mass
$m_1=m_2$. This approximation is supported by the observed
differences in the squares of neutrino masses as
measured by oscillations of atmospheric
neutrinos, $|\Delta m_{23}^2|/\Delta
m_{12}^2\simeq0.5\cdot10^{-2}$ \cite{Maltoni:2004ei}. The advantage
of this parameterization is that it allows us to represent in a simple
way various mass hierarchies such as total degeneracy ($\alpha=1/3$)
and normal or inverted hierarchy given respectively by $\alpha\sim1$
and $\alpha\ll1$.

\section{Forecast for Weak Lensing Tomography}\label{forecast}
Weak lensing (see \cite{art:Munshi} for a recent
review or {\tt http://www.gravitationallensing.net}) is a
particularly powerful probe of cosmology since it simultaneously
measures the growth of structure through the matter power spectrum,
and the geometry of the Universe through the lensing effect. Since
weak lensing probes the dark matter power spectrum directly it is
not limited by assumptions about galaxy bias (how galaxies are
clustered with respect to the dark matter). Future weak lensing
surveys will measure photometric redshifts of billions of
galaxies allowing the
possibility of 3D weak lensing analysis (e.g. \cite{Heavens:2003,Castro:2005,Heavens:2006,Kitching:2007}) or a tomographic reconstruction of growth of structures
as a function of time
through a binning of the redshift distribution of galaxies $D(z)$,
with a considerable gain of cosmological information (e.g. on
neutrinos \cite{Hannestad:2006as}; dark energy
\cite{Kitching:2007}; the growth of structure
\cite{art:Massey,art:Bacon}  and map the dark matter distribution as
a function of redshift \cite{art:Taylor}).

In this section we present forecasts for the upcoming weak lensing
survey Euclid \cite{Refregier:2006vt}, combined with constraints
expected from the CMB Planck experiment \cite{:2006uk} using a Fisher matrix
formalism. We first explored a simple $8$-parameter model with
fiducial values $\Omega_bh^2=0.022$, $\Omega_ch^2=0.111$, $h=0.7$,
$\tau=0.084$, $n_s=0.95$, $As=2.48\cdot10^{-9}$, $\sum
m_{\nu}=0.055{\rm eV}$ and the hierarchy parameter $\alpha=0.95$. Note
that we are assuming a fiducial normal hierarchy scheme for neutrino
masses. We then repeated the analysis for a larger set of
parameters, including running of spectral index (with target value
$dn_s/d\ln k=0$) and the dark energy equation of state $w$ (assuming
cosmological constant $w=-1$ as the fiducial model), to study how degeneracies
with various parameters affect constraints on neutrino mass
hierarchy.

To calculate the power spectra we used the numerical code
CAMB \cite{Lewis:1999bs} that can calculate CMB and matter power
spectra, also for non-degenerate neutrino masses. The usual
definition of the Fisher matrix \cite{art:Tegmark} is:
\begin{equation}\label{fisher}
F_{\alpha\beta}\equiv\left\langle-\frac{\partial^{2}\ln L}{\partial
p_{\alpha}\partial p_{\beta}}\right\rangle
\end{equation}
where $L$ is the likelihood function for a set of parameters $p_i$.
When the derivatives of (\ref{fisher}) are evaluated at the fiducial
model the Fisher matrix
gives an estimate of the best statistical error achievable on the
parameters (via the Cramer-Rao inequality) for the method and survey
design considerd $$\sigma(p_{i})\ge
\sqrt{(F^{-1})_{ii}}.$$
The Fisher matrix for weak lensing is given by
(e.g. \cite{Amendola:2007rr})
\begin{equation}\label{fisherwl}
    F^{WL}_{\alpha\beta}=f_{sky}\sum_{\ell}\frac{(2\ell+1)\Delta\ell}{2}
    \frac{\partial
    P_{ij}}{\partial p_{\alpha}}C_{jk}^{-1}\frac{\partial
    P_{km}}{\partial p_{\beta}}C_{mi}^{-1}
\end{equation}
where $P_{ij}(\ell)$ is the convergence weak lensing power spectrum
that depends on the non-linear matter power spectrum at redshift
$z$, $P_{nl}(k,z)$, obtained by correcting the linear matter power
spectrum $P(k,z)$ using the option $\texttt{halofit}$ of CAMB
\cite{Smith:2002dz}. $\Delta\ell$ is the step used for $\ell$ and
\begin{equation}\label{c}
    C_{jk}=P_{jk}+\delta_{jk}\langle\gamma^2_{int}\rangle n_j^{-1}.
\end{equation}
In the last expression $\gamma_{int}$ is the rms intrinsic galaxy ellipticity
(and we assume $\langle\gamma^2_{int}\rangle^{1/2}=0.22$) and $n_j$
is the number of galaxies per steradian belonging to $j^{\rm th}$ bin
\begin{equation}\label{number}
    n_j=3600d\left(\frac{180}{\pi}\right)^2\hat{n}_{j},
\end{equation}
where $d$ is the number of galaxies per square arcminute and
$\hat{n}_j$ is the fraction of sources belonging to the $j^{\rm th}$
bin. For the Euclid experiment we take $d=35$ and $f_{sky}=0.5$. The
galaxy redshift distribution is assumed to have the form $D(z)
\propto z^2\exp[-(z/z_0)^{1.5}]$ with $z_0=0.9$. For this experiment
photometric redshift uncertainties are assumed to be
$\sigma_z=0.03(1+z)$. We treat this uncertainty following the
approach of \cite{Ma:2005rc} where a galaxy with redshift $z$ could
be wrongly observed at a redshift $z_{ph}$. Letting $p(z|z_{ph})$ be
the probability that this happens the distribution of galaxies in
the $i^{\rm th}$ bin is modified to take into account this `leakage'
between bins $D_{i}(z)=\int_{z_{ph,i}}dz_{ph}D(z)p(z|z_{ph})$. We
choose a simple Gaussian form for $p(z|z_{ph})$
$$p(z|z_{ph})=(2\pi\sigma_z^2)^{-1/2}\exp\left[-\frac{(z_{ph}-z)^2}{2\sigma_z^2}\right].$$

The Fisher matrix for weak lensing is then added to that of CMB to
obtain constraints from the combination Planck$+$Euclid $$F^{tot}_{\alpha\beta}=F^{WL}_{\alpha\beta}+F^{CMB}_{\alpha\beta}.$$
For a CMB experiment the Fisher matrix is given by \cite{Bond:1997wr}:
\begin{equation}\label{fishcmb}
    F_{\alpha\beta}^{CMB}=\sum_{\ell=2}^{\ell_{\rm max}}\sum_{PP',QQ'}\frac{\partial
    C_{\ell}^{PP'}}{\partial p_{\alpha}}(Cov_{\ell}^{-1})_{PP',QQ'}\frac{\partial
    C_{\ell}^{QQ'}}{\partial p_{\beta}}
\end{equation}
where the couples $PP'$ and $QQ'$ mean in our case $TT$, $TE$ or
$EE$ (temperature and E-mode polarisation).
$Cov_{\ell}$ is the power spectrum covariance matrix at the
$\ell^{\rm th}$ multipole and $\ell_{\rm max}$ is the maximum multipole available
given the angular resolution of the considered experiment, for
Planck we use $\ell_{\rm max}=2000$. The other specifications we used
are listed in Table \ref{Planck}. The total Fisher matrix $F^{tot}$
is then inverted to obtain uncerainties on cosmological parameters
of our model.
\begin{table}[h]
\begin{center}
\begin{tabular}{rcccc}
\hline
PLANCK& Channel/GHz & FWHM & $\Delta T/T$ & $\Delta P/T$  \\
\hline
$f_{\rm sky}=0.65$
& 100 & $9.5'$ & 2.5 & 4.0 \\
& 143 & $7.1'$ & 2.2 & 4.2 \\
& 217 & $5.0'$ & 4.8 & 9.8 \\
\hline
\end{tabular}
\caption{specifications for Planck experiment used in the Fisher
matrix calculation. $\Delta T/T$ and $\Delta P/T$ are sensitivities
($\mu K/K$) for temperature and polarization respectively.}
\label{Planck}
\end{center}
\end{table}

\section{Results}\label{results}
In this section we show results of our forecasts on
 neutrino mass parameters $\sum m_{\nu}$
and $\alpha$.

\subsection{Parameter Constraints}

\begin{figure}[h]
\begin{center}
\includegraphics[width=250pt]{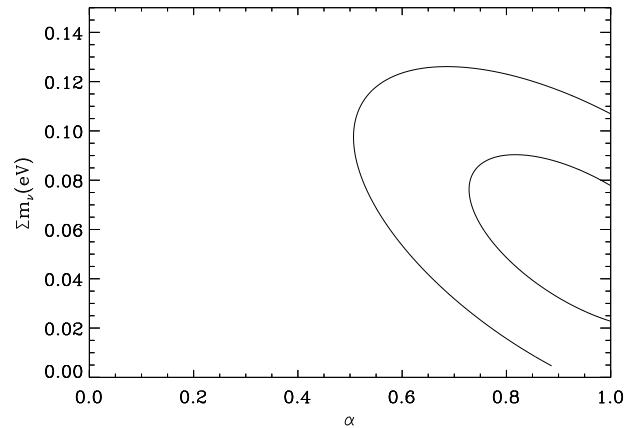}
\caption{$68\%$ and $95\%$ probability contours (two-parameter) in the plane
$\alpha$-$\sum m_{\nu}$ for Planck+Euclid from our Fisher matrix
calculation for the $10$-parameter space described in the text.}\label{contour1}
\end{center}
\end{figure}

The constraints of Table \ref{sigmas} and Fig. \ref{contour1} show
that for $\sum m_{\nu}=0.055{\rm eV}$, if the neutrino mass
hierarchy is described by a normal hierarchy ($\alpha=0.95$) then
the combination of weak lensing data from an experiment like Euclid
and accurate measurements of the CMB power spectrum, achievable with
Planck, can become sensitive to the mass of single species. The
$1\sigma$ uncertainties on the sum of masses is $\sigma_{\sum
m_{\nu}}=0.037{\rm eV}$, which is in agreement with other forecasts
(see for example \cite{Kitching:2008dp}) and confirms the ability of
these surveys to detect neutrino mass. In particular this work has a
resonance with Debono et al. (2009) (in preparation) in which the
effect of parameter sets on neutrino mass constraints are
investigated. We find that for the parameters that are common
between the two articles there is agreement between the predicted
errors, despite the slightly different parameter sets, power
spestrum estimation approaches (CAMB vs Eisenstein \& Hu code
\cite{EH}) and assumptions.

As one can see for a typical normal hierarchy scenario with
$\alpha=0.95$ the combination Euclid+Planck can reach a $\sim20\%$
sensitivity on $\alpha$ with an error $\sigma_{\alpha}=0.19$. For
this normal hierarchy scenario we have repeated the Fisher matrix
calculation using a larger parameter space, including running $d
n_s/d\ln k$ of the spectral index and the dark energy equation of
state parameter $w$, to check the weakening in the constraints on
$\alpha$ induced by degeneracies among these parameters. One may
expect that running and $w$ would have a large degeneracy with
neutrino mass since they both can add an effective damping on small
scale. However, as shown in in Table \ref{sigmas} and in Fig.
\ref{contour1}, constraints on $\alpha$ are not seriously weakened
-- for this second $10$-parameter space $\sigma_{\alpha}=0.22$.
\begin{table}[h]
\begin{center}
\begin{tabular}{lccc}
experiment  &&  $\sigma_{\sum m_{\nu}}$ & $\sigma_{\alpha}$\\
\hline
Planck      &&      $0.49$ {\rm eV}  &  $>1$ \\
Planck+Euclid &&  $0.037$ {\rm eV} & $0.22$\\
\hline
\end{tabular}
\end{center}
\caption{contraints from Planck+Euclid Fisher matrix on neutrino
mass parameters
for the $10$-parameter space described in the text.}\label{sigmas}
\end{table}

\subsection{Parameterisation Investigation}

\begin{figure}
\begin{center}
\includegraphics[width=250pt]{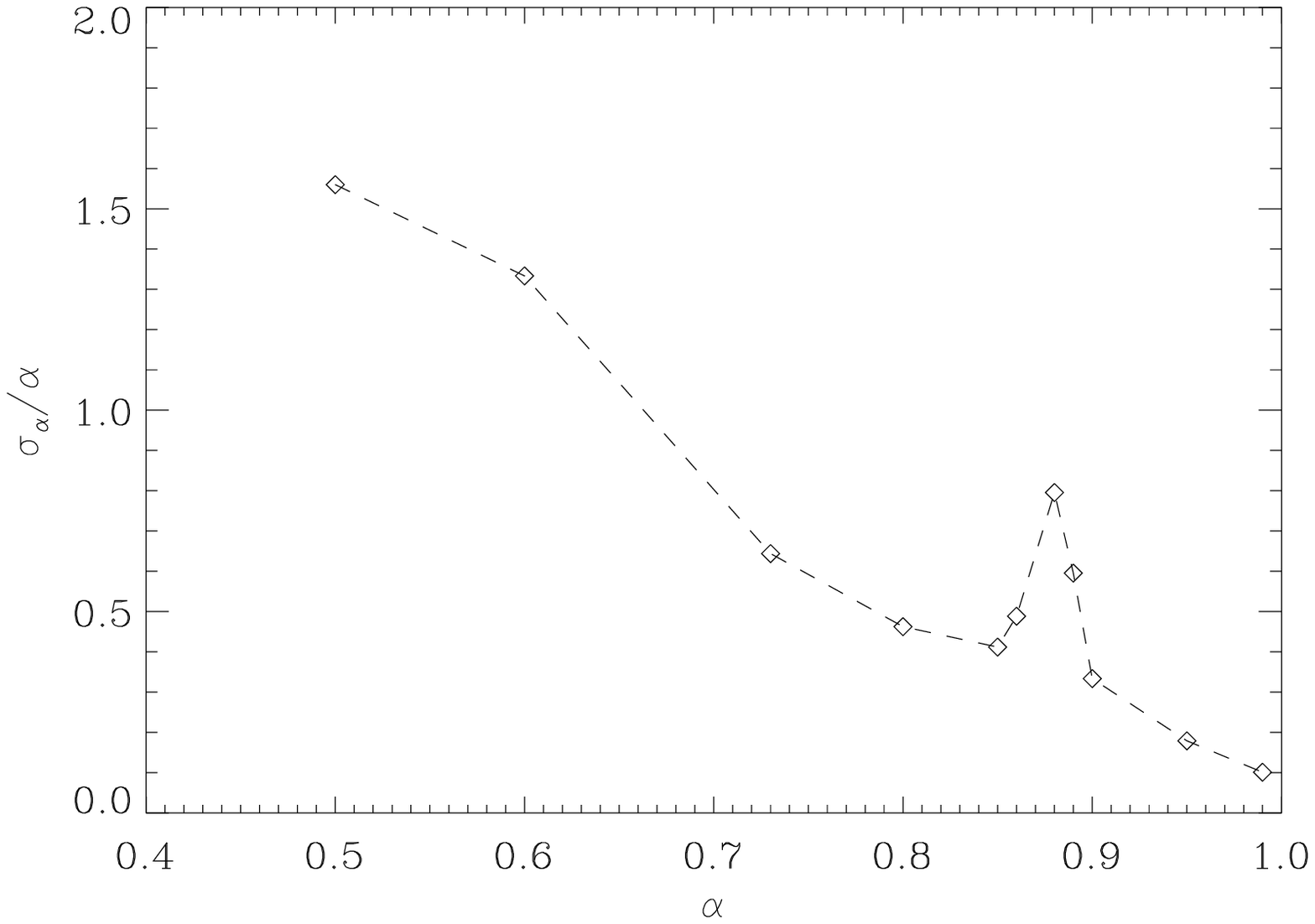}
\caption{relative error on $\alpha$ as a function of the target model for the combination Planck$+$Euclid. The fiducial value for the total mass is $\sum m_{\nu}=0.055${\rm eV}.}\label{sigalpha}
\end{center}
\begin{center}
\includegraphics[width=\columnwidth,clip=true]{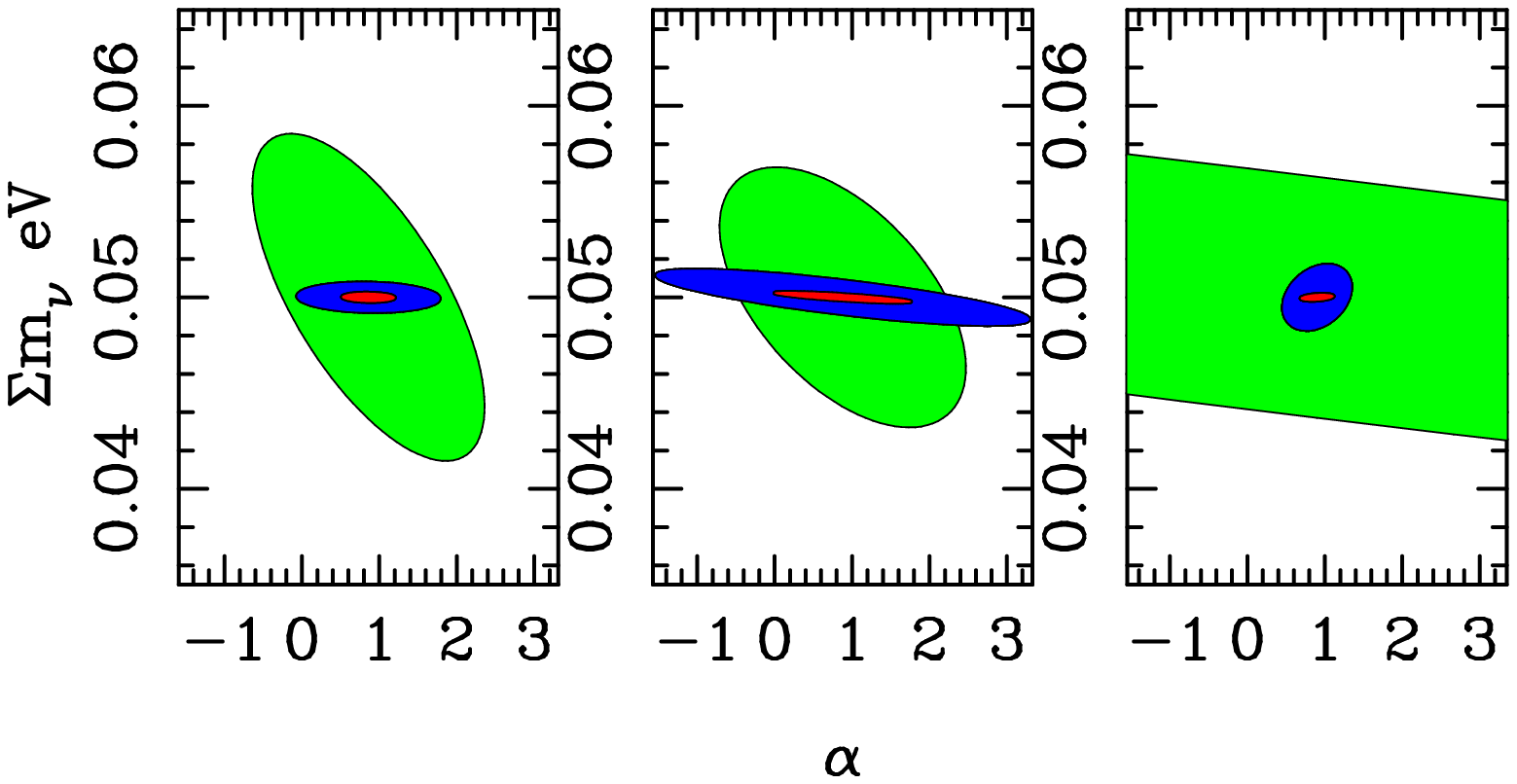}
\caption{$68\%$ $2$-parameter probability contours in the plane
$\sum m_{\nu}$-$\alpha$ for Planck (green, light ellipse), Euclid
  (blue, dark ellipse) and the combination (red, central ellipse)
  from our Fisher matrix
  calculations for the $8$-parameter space described in the text. The
  plots are for fiducial values $\alpha=0.86$ (left), $\alpha=0.88$
  (middle) and $\alpha=0.9$ (right). }\label{ellipses}
\end{center}
\begin{center}
\includegraphics[width=250pt]{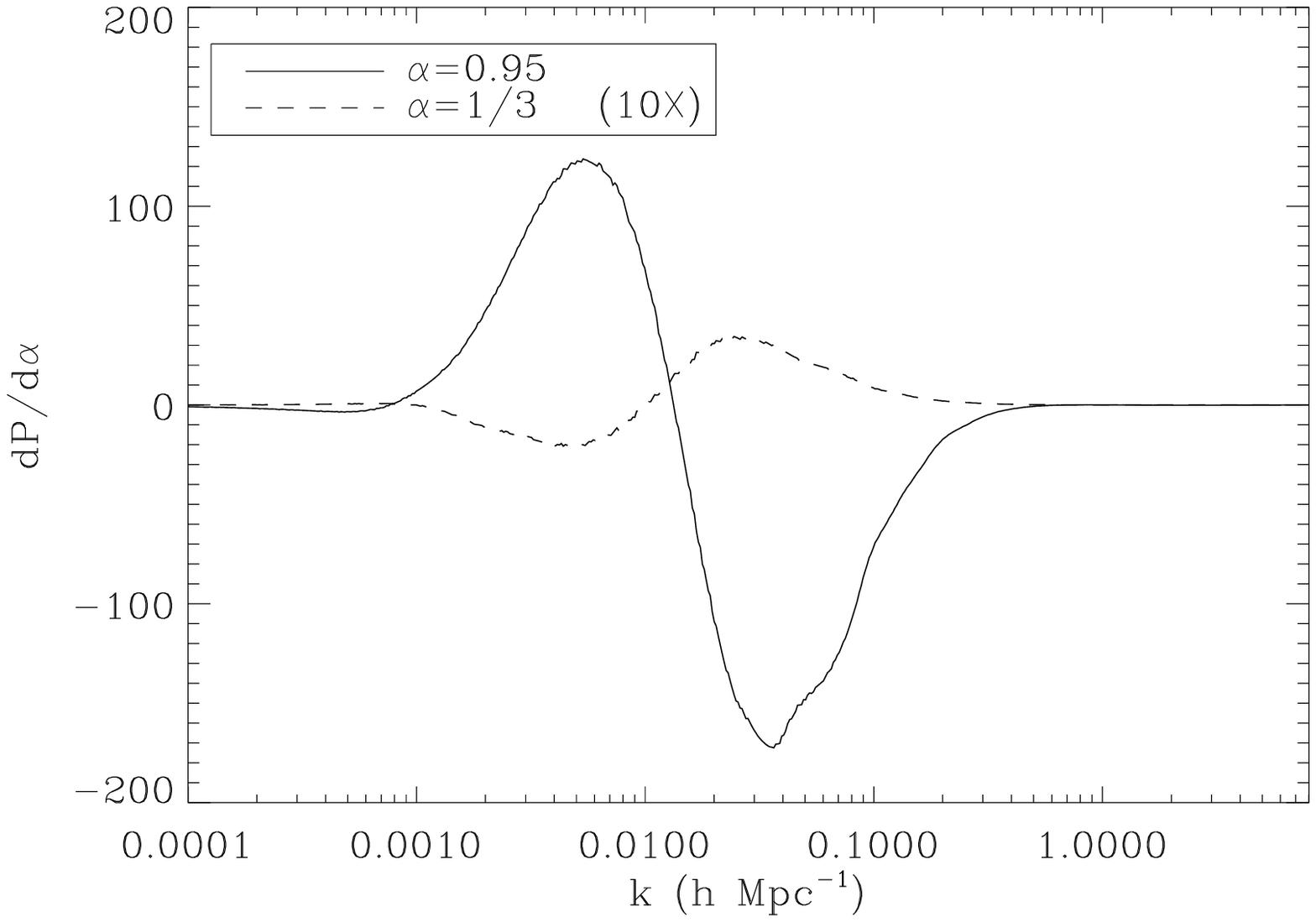}
\caption{derivatives of matter power spectrum with respect to $\alpha$
  for two target models. The derivative for $\alpha=1/3$ is multiplied
  for a factor $10$. See text for more details.}\label{der}
\end{center}
\end{figure}

For the smaller $8$-parameter set we repeated the Fisher matrix calculation
for different fiducial values for $\alpha$. In Fig. \ref{sigalpha}
we show the relative uncertainties in $\alpha$ from the
combination Euclid$+$Planck as a function of the
target model. The figure is a combination of a general trend, that
causes the uncertainty on $\alpha$ to decrease for $\alpha$ increasing,
and the way in which constraints from CMB combine with those from
weak lensing. The loss of sensitivity around $\alpha\simeq0.88$ is in
fact due to a strong rotation of the degeneracy in the plane $\alpha$-$\sum
m_{\nu}$ in the Euclid Fisher matrix, as shown in
Fig. \ref{ellipses}, from an anti-correlation to a positive correlation. For
the Planck Fisher matrix instead the degeneracy between these two
parameters is less dependent on the fiducial value of $\alpha$. As one
can see for $\alpha\simeq0.88$ the constraints from Euclid are almost
independent of $\alpha$, and the combination of the two
experiments lose the ability to break the degeneracy between these
parameters. This is confirmed also by the fact that the only other
parameter that shows a significant \textit{peak} of uncertainty in
correspondence of $\alpha\simeq0.88$ is the total mass $\sum m_{\nu}$
itself that icrease from $\sigma_{\sum m_{\nu}}\simeq0.025{\rm eV}$
for $\alpha=0.86$ to $\sigma_{\sum m_{\nu}}\simeq0.034{\rm eV}$ for
$\alpha=0.88$.

We emphasise that this is a peculiar effect of this parameterisation,
we have tested the derivative numerically for convergence. Therefore
we advocate this parameterisation but with a strong warning that
results are highly dependant on the fiducial value of $\alpha$,
particularly around $\alpha\approx 0.88$.

We note however that the general decrease in $\sigma_{\alpha}/\alpha$
is quite intuitive since for $m_3\simeq\sum m_{\nu}$
a variation in $\alpha$ means a variation only in the mass that can
have an effect on the growth of structure. In this limit the others
two eigenstates have a very low mass and are relativistic up to a very
low redshift. For example for our target model ($\sum
m_{\nu}\simeq0.055{\rm eV}$ and $\alpha=0.95$) the two eigenstates have
$m_1=m_2\simeq1.4\cdot10^{-3}{\rm eV}$, this results in the neutrinos being
relativisitc up to a redshift $z\sim2$. Hence for
$\alpha\rightarrow1$ the sensitivity to this parameter increases.
Conversely when $\alpha$ is significantly different from $1$, for
example in the case of total degeneracy, when $\alpha=1/3$, a
variation in $\alpha$ and hence in $m_3$ implies an opposite
variation in $m_1$ and $m_2$ which now have a non-negligible mass.
These variations partially compensate reducing the sensitivity of
cosmology to $\alpha$.

This reasoning is confirmed by
the comparison between derivatives of the matter power spectrum
$P(k)$ (that enters in the calculation of the weak lensing
convergence power spectrum) with respect to $\alpha$ calculated for
different values of $\alpha$. In Fig. \ref{der} are shown
derivatives $dP/d\alpha$ for the case $\alpha=0.95$ and
$\alpha=1/3$. The latter is about two orders of
magnitude smaller than the first. As one can see the two derivatives
have opposite signs due to the different change in $k_{FS}$
induced by a change in $\alpha$ in the two target models. When
$\alpha\rightarrow1$ an increase in $\alpha$ causes an increase
in $k_{FS}$ according to equation (\ref{kfs}) while, as we have said above,
$m_1$ and $m_2$ are not important. Instead, for $\alpha=1/3$ an
increase in this parameter causes a decrease of $m_1$ and $m_2$ (which
now cannot be neglected) and the overall effect is a decrease in
$k_{FS}$.

\section{Bayesian analysis}\label{Bayes}
To assess better the power of these cosmological probes to
detect a neutrino mass difference we present a Bayesian evidence
forecast. Calculation of Bayesian evidence
allows one to make a comparison of different \emph{models}, as opposed
to parameter estimation \emph{within} a model. According to
Bayes theorem the (posterior) probability of a set of parameters
$\theta$ describing a model $\mu$ given the data $d$ is
\cite{Trotta:2008qt}
\begin{equation}\label{bayes}
p(\theta|d,\mu)=\frac{\mathcal{L}(\theta)p(\theta|\mu)}{p(d|\mu)},
\end{equation}
where $\mathcal{L}(\theta)=p(d|\theta,\mu)$ is the likelihood
function,
$p(\theta|\mu)$ is the prior probability on the parameters of
the model $\mu$ and $p(d|\mu)$ is the Bayesian evidence; that has the
role of a normalization constant, being
 $p(d|\mu)=\int d\theta p(d|\theta,\mu)p(\theta|\mu)$.

The following is a summary of the technique describe in \cite{Heavens:2007ka}.
Given a cosmological model $M'$ described by a number of parameters
$n'$,
a common problem is to verify whether data require the
inclusion of some new parameters in the model so as creating a new
more complicated model $M$ with a number of parameter $n>n'$.
In this case one has to take the ratio of the posterior probabilities
of the two models $p(M'|d)/p(M|d)$. This ratio can be obtained
from Bayes thorem:
\begin{equation}\label{ratio}
\frac{p(M'|d)}{p(M|d)}=B\frac{p(M')}{p(M)}
\end{equation}
where $p(M')$ and $p(M)$ are the prior probabilities of the two models
and $B$ is the \textit{Bayes factor} given by the ratio
between Bayesian evidences:
\begin{equation}\label{bayesfactor}
B=\frac{\int d\theta' p(d|\theta',M')p(\theta'|M')}{\int d\theta
  p(d|\theta,M)p(\theta|M)}
\end{equation}
where $\theta'$ and $\theta$ are the set of parameters of $M'$ and
$M$. In the case we are considering the two models are
\textit{nested},
 in the sense that they share the same $n'$ parameters. For nested
 models and for Gaussian likelihoods (\ref{bayesfactor}) approximates to
\cite{Heavens:2007ka}
\begin{equation}\label{bayesfactor2}
B=(2\pi)^{-p/2}\frac{\sqrt{\det F}}{\sqrt{\det F'}}\exp\left(-\frac{1}{2}\delta\theta_{\alpha}F_{\alpha\beta}\delta\theta_\beta\right)\prod_{q=1}^{p}\Delta\theta_{n'+q},
\end{equation}
where the index $q$ runs over the $p=n-n'$ additional parameters of
$M$ with respect to $M'$ and  $\Delta\theta$ are the prior
ranges on the parameters. Note that the Fisher matrix $F$ is $n\times
n$ while $F'$ is $n'\times n'$.

The $p$ parameters are assumed
to be fixed at a certain fiducial values in the model $M'$ are shifted
by  an amount $\delta\psi$ with respect to their fiducial value in $M$.
For $\alpha,\beta=1,...,n'$ then this shift $\delta\theta_{\alpha}$ is given by
\begin{equation}\label{delta}
\delta\theta_{\alpha}=-(F'^{-1})_{\alpha\beta}G_{\beta\gamma}\delta\psi_{\gamma}\hspace{1cm}\gamma=1,...,p,
\end{equation}
where for $\alpha,\beta=1,...,p$ we have
$\delta\theta_{\alpha}=\delta\psi_{\alpha}$. The quantity $G$ that
appears in (\ref{delta}) is a block $n'\times p$ of the full Fisher
matrix $F$.

The calculation of Bayes factor through Fisher matrices
helps to clarify whether future experimental data will be sensitive
to a wrong assumption about some parameters (for example fixing
$p$ parameters to wrong fiducial values). Bayesian analysis is
known to be \textit{conservative} in the sense that models with a
smaller number of parameters are favoured until data strongly
require the introduction of new parameters \cite{Trotta:2008qt}.
Hence, for the case of nested models only very sensitive experiments
will have the power to discern that a certain parameter is kept fixed
to an incorrect value.

In what follows we apply equation (\ref{bayesfactor2}) to our neutrino
Fisher matrices so that we can
understand if future weak lensing and CMB data will achieve enough
sensitivity to require a parameterization of the
mass hierarchy. All results showed below are for the combination of
the Euclid and Planck experiments and
hence the Fisher matrix used in the calculation of Bayes factor is the
total Fisher matrix
$F^{tot}_{\alpha\beta}=F_{\alpha\beta}^{WL}+F^{CMB}_{\alpha\beta}$. We
assume that the \textit{true} model $M$ is
represented by the $10$-parameter model described in the previous
section with a normal hierarchy scheme for neutrinos ($\alpha=0.95$)
and $\sum m_{\nu}=0.055{\rm eV}$.  We next consider a simpler $9$-parameter
model $M'$ in which the parameter $\alpha$ is fixed to a certain
value, shifted of an amount $\delta\alpha$ with respect to
$\alpha=0.95$ and calculate the Bayes factor for these two competing
models.
\begin{figure}[h]
\begin{center}
\includegraphics[width=250pt]{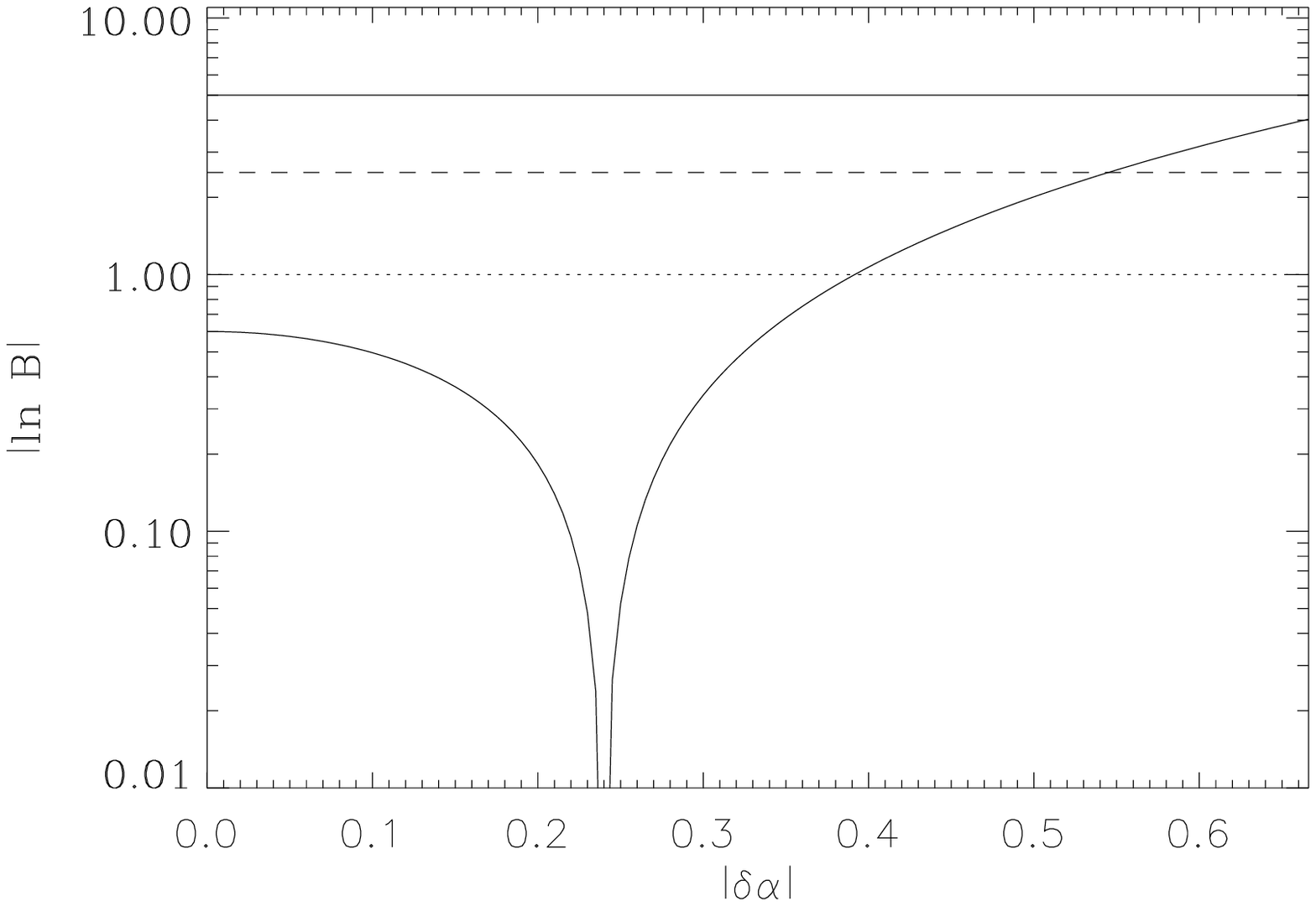}
\caption{absolute value of $\ln B$ as a function of $|\delta\alpha|$.
The lines indicates the limits of the Jeffreys scale. On the right of the cusp is $B<1$, meaning evidence favours
a more general parameterization of neutrino mass hierarchy.}\label{lnb}
\includegraphics[width=250pt]{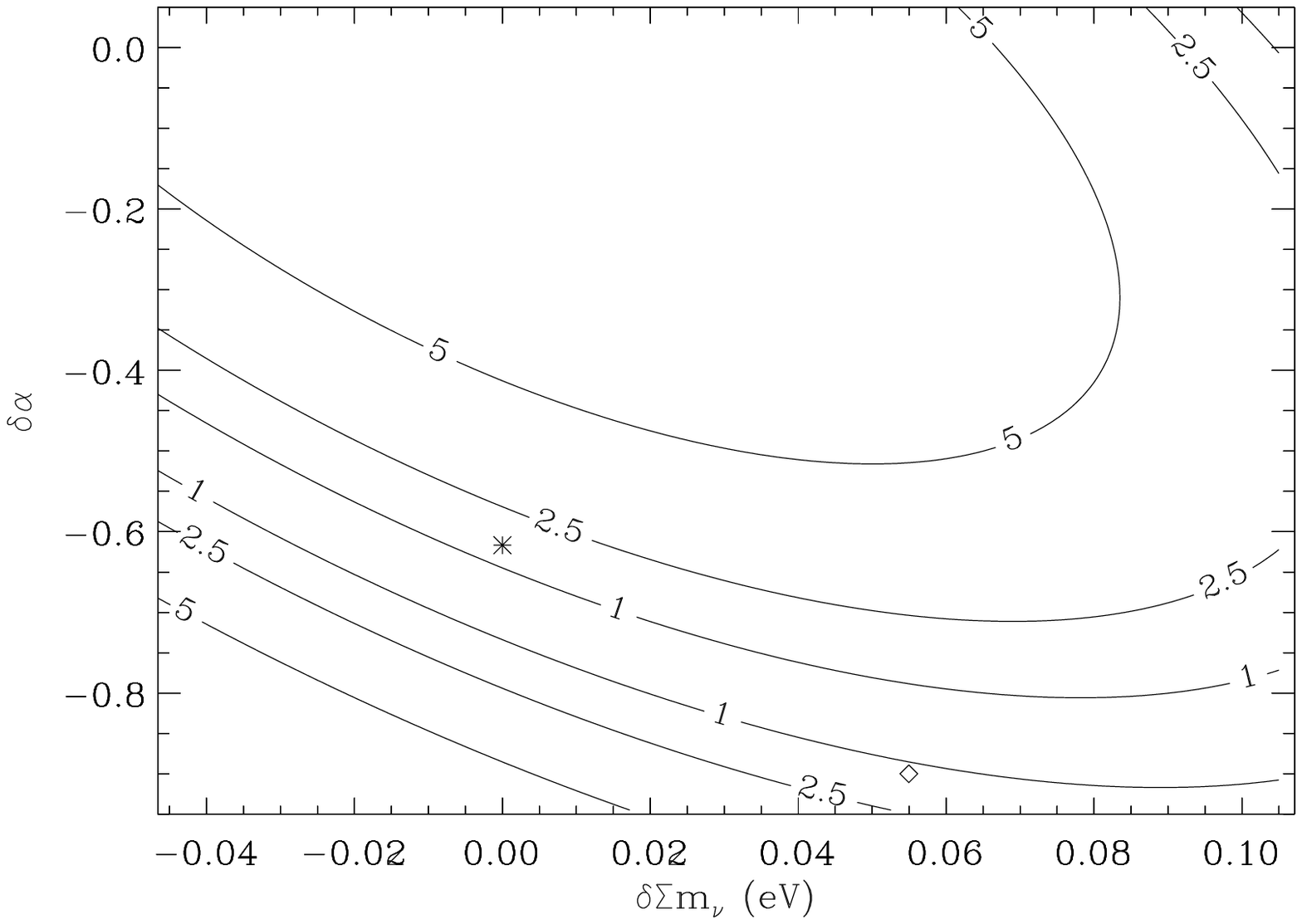}
\caption{Jeffreys scale contours of $\ln B$ as a function of $\delta\alpha$ and $\delta\sum m_{\nu}$. The inner part of the plot (from the innermost $\ln B=1$ contour) corresponds to values $B>1$ and hence evidence for the simpler model (see text). The cross indicates the assumption of degenerate masses fixing the total mass to the correct value; the square indicates a typical inverted hierarchy scenario.}\label{contourlnb}
\end{center}
\end{figure}

Results of the evidence calculation are shown in Fig. \ref{lnb} where we
plot $|\ln B|$ as a function of $|\delta\alpha|$. The horizontal
lines corresponds to the values of the Jeffreys scale
\cite{Jeffreys}: $|\ln B|<1$ means \textit{inconclusive} evidence,
$1<|\ln B|<2.5$ is a \textit{substantial} evidence, $2.5<|\ln B|<5$
is a \textit{strong} evidence and $|\ln B|>5$ is considered
\textit{decisive} evidence. The area leftward of the
cusp corresponds to a Bayes factor $B>1$ which in our case means
evidence for the simpler model $M'$. For
$|\delta\alpha|\gtrsim0.24$, $B$ becomes smaller than $1$ indicating
that data would require the introduction of $\alpha$ in the analysis,
and will be sensitive to differences between neutrino masses. In
particular for $\delta\alpha\simeq-0.62$, that represents the common
(and under our hypothesis \textit{wrong}) assumption of total
degeneracy between neutrino masses, the data would give a strong
evidence for model $M$ requiring a parameterization of neutrino
mass differences.

If data will be able to give evidence for a
hierarchy of neutrino masses then it is proper to verify what
the effect of assuming degeneracy
of masses, or a wrong hierarchy, on other cosmological parameters will
be. As
shown in \cite{Heavens:2007ka}
fixing one parameter to a wrong value causes a shift in the best fit
value of other parameters according to (\ref{delta}). We calculate
the bias in cosmological parameters due to a wrong assumption for
$\alpha$ and assuming normal hierarchy ($\alpha=0.95$) as true
model. The results are shown in Table \ref{shifts} and show that
assuming a degenerate hierarchy ($\delta\alpha=-0.62$) or an inverted
hierarchy ($\delta\alpha=-0.9$) would cause a shift in other
cosmological parameters comparable with the $1\sigma$ statistical error.
Only the shift in $\sum m_{\nu}$ is significantly greater than
the $1\sigma$ error due to the high degeneracy between
$\alpha$ and $\sum m_{\nu}$.

These results
confirm those of Fig. \ref{lnb}; the data will be accurate enough to require
a parameterization of the mass splitting, this causes a non-negligible bias
in other parameters in the case of a wrong assumption.
However the bias on cosmological
parameters is generally smaller than the $1\sigma$ uncertainties. Notable
exceptions include the dark energy equation of state parameter.
Assuming total degeneracy of neutrino masses, the shift on $w$ is
smaller than $1\sigma$ but becomes slightly greater  assuming an
inverted hierarchy. Our Fisher matrix analysis indicates that this is
due essentially to the shift in $\sum m_{\nu}$ and hence to the
degeneracy $w$-$\sum m_{\nu}$ because there is no significant
correlation between $w$ and $\alpha$.
\begin{table}
  \centering
\begin{tabular}{l|l|l|ll}
&&Inverted&Degenerate\\
\hline
Parameter&$\sigma$&$\delta\alpha=-0.90$&$\delta\alpha=-0.62$\\
  \hline
$w$&$0.041$&             $-0.047$&                          $-0.033$\\
  $\Omega_bh^2$ &          $10^{-4}$ &   $-0.4\cdot10^{-4}$&   $-2.7\cdot10^{-5}$  \\
  $\Omega_ch^2$ &          $0.00065$ &   $0.0013$ &           $0.00090$\\
  $h$ &                  $0.013$ &      $0.0049$ &             $0.0036$\\
  $\tau$ &                 $0.0028$&     $-0.0012$&           $-0.00082$ &\\
  $n_s$ &                  $0.0022$&     $-0.0036$&           $-0.0024$ &\\
  $A_s$ &                  $1.44\cdot10^{-11}$&$5.75\cdot10^{-12}$               &$3.94\cdot10^{-12}$& \\
  $\sum m_\nu ({\rm eV})$ &           $0.037$ &         $0.086$            &    $0.060$ &\\
  $dn_s/d\ln k$ &          $0.0031$ &          $-0.0019$                      & $-0.0012$ &\\\end{tabular}
   \caption{$1\sigma$ errors on cosmological parameters and the bias
($\delta$) due to a wrong assumption in the neutrino mass
hierarchy.}\label{shifts}
\end{table}

We also show the results of an evidence calculation for an even simpler model
$M''$, in which both the total mass $\sum m_{\nu}$ and $\alpha$ are
fixed, over the model $M$. Note that now we are comparing an
$8$-parameter model with a $10$-parameter model. The results are shown
in Fig. \ref{contourlnb} where we have plotted the contour values of
$|\ln B|$ as a function of $\delta{\alpha}$ and $\delta{\sum
m_{\nu}}$. The inner contours from $1$ to $5$ correspond to values
$B>1$, and so to an evidence which favours the simpler model, while the outer
contours are relative to values $B<1$ and hence to evidence for the
more complicated model $M$. The star in the plot refers to the wrong
assumption of total degeneracy (but fixing the total mass to the true
value $\delta{\sum m_{\nu}}=0$).
As one can see, under these
assumptions the data would favour the simpler model giving a substantial
evidence for $M''$.

The evidence for the simpler model is due to
the smaller number of parameters of $M''$ with respect to $M$ (the
Occams razor term, see \cite{Heavens:2007ka}) and also to the
negative degeneracy between $\alpha$ and $\sum m_{\nu}$ (given by
the off-diagonal term of the inverted full Fisher matrix $(F^{-1})_{\sum
m_{\nu}\alpha}/\sqrt{(F^{-1})_{\alpha\alpha}(F^{-1})_{\sum
m_{\nu}\sum m_{\nu}}}\simeq-0.44$).  Because of this degeneracy there
is a \textit{region of confusion} in the plane of Fig.
\ref{contourlnb} in which a wrong assumption in $\alpha$ is
compensated by a wrong, and opposite in sign, assumption for $\sum
m_{\nu}$, leading to evidence in favour of the simpler model $M''$.
This explains why in Fig. \ref{contourlnb} for a fixed
$\delta\alpha$ the the Bayes factor initially increases becoming
greater than $1$ for $\delta \sum m_{\nu}$ increasing. Of course
when $\delta \sum m_{\nu}$ becomes large enough the Bayes factor
decreases because the data start to favour the true model $M$.

If one assumes
an inverted hierarchy (represented by the square in the plot), for
example fixing the total mass to the minimum value allowed for this
hierarchy ($\sum m_{\nu}\simeq0.11{\rm eV}$) and $\alpha$ to a low value
(we take $\alpha=0.05$), one obtains a strong evidence for the more
complicated model $M$.

These evidence calculations indicate that a future weak lensing
survey could become sensitive to the hierarchy of neutrino masses
requiring a suitable parameterization of mass splitting. We note
however that even if a model assuming total degeneracy could
be still favoured. This indicates that
Planck and Euclid data could give strong evidence for the
existence of a hierarchy among neutrino masses.

\section{Conclusions}\label{conclusions}

In this paper we have investigated the ability of future cosmic shear
measurements, like those achievable with the proposed Euclid mission,
 to constrain differences in the mass of individual neutrino species.
Using an explicit parameterization of neutrino mass splitting and a Fisher
matrix formalism we have found that the combination Euclid+Planck
will reach enough sensitivity to put constraints on the fraction of
mass in the third neutrino mass eigenstate ($\alpha$) under the
assumption that the neutrino mass scheme is described by a normal
hierarchy, with fiducial value $\alpha=0.95$:
$\sigma_{\alpha}=0.22$. We have also investigated the
parameterization dependence of our results, repeating our forecasts
calculation for different fiducial values of $\alpha$ and finding
that constraints on $\alpha$ are generally decreasing for increasing
$\alpha$. We have found a loss of sensitivity around
$\alpha\simeq0.88$ due to a strong rotation of the degeneracy in the
plane $\sum m_{\nu}$-$\alpha$ in the Euclid Fisher matrix,
suggesting a considerable dependence of the results on the fiducial
value around this point.\\ We have then studied more deeply
 the power of detecting neutrino mass splitting
considering Bayesian evidence. We have found that future data from
these experiments will provide strong evidence for a neutrino mass
splitting against a model assuming degeneracy of masses. As a
consequence we have also shown that assuming a wrong hierarchy can
bias constraints on other cosmological parameters, in particular
with those parameters involved in degeneracies with $\sum m_{\nu}$
and $\alpha$. For the dark energy equation of state we found a bias
greater than the $1\sigma$ statistical error ($\sigma_w=0.041$ from Euclid+Planck)
assuming an inverted hierarchy $\delta_w=-0.047$ and comparable with
$1\sigma$ assuming degenerate masses $\delta_w=-0.033$.\\ In
conclusion we emphasize that, even if these constraints are strongly
dependent on the parameterization used, the possibility of having a
splitting in neutrino masses cannot be neglected in analysing
future data from Euclid-like experiments.  A wrong assumption about
neutrino mass hierarchies can indeed cause non-negligible bias on
other cosmological parameters and in particular on the dark energy
equation of state.\\\\
\textit{Acknowledgements}\\\\
TDK is supported by STFC rolling grant number RA0888. We thank Adam
Amara, Asantha Cooray, Julien Lesgourgues, Luca Pagano, Paolo Serra
and An\v{z}e Slosar for useful discussions. FDB thanks Institute for
Astronomy and Royal Observatory of Edinburgh for hospitality while
this research was conducted.

\end{document}